\documentclass[pre]{revtex4-2}

\usepackage{amssymb, mathtools, siunitx}
\usepackage[dvipdfmx]{graphicx}

\DeclareMathOperator{\erfc}{erfc} 

\begin{document}
\title{Theory and data analysis of player and team ball possession time in football}
\author{Ken Yamamoto}
\affiliation{Faculty of Science, University of the Ryukyus, Nishihara, Okinawa 903-0213, Japan}
\author{Seiya Uezu}
\affiliation{Graduate School of Engineering and Science, University of the Ryukyus, Nishihara, Okinawa 903-0213, Japan}
\author{Keiichiro Kagawa}
\affiliation{Research Institute for Electronic Science, Hokkaido University, Sapporo, Hokkaido 060-0812, Japan}
\author{Yoshihiro Yamazaki}
\affiliation{School of Advanced Science and Engineering, Waseda University, Shinjuku, Tokyo 169-8555, Japan}
\author{Takuma Narizuka}
\affiliation{Faculty of Data Science, Rissho University, Kumagaya, Saitama 360-0194, Japan}

\begin{abstract}
In this study, the stochastic properties of player and team ball possession times in professional football matches are examined.
Data analysis shows that player possession time follows a gamma distribution and the player count of a team possession event follows a mixture of two geometric distributions.
We propose a formula for expressing team possession time in terms of player possession time and player count in a team's possession, verifying its validity through data analysis.
Furthermore, we calculate an approximate form of the distribution of team possession time, and study its asymptotic property.
\end{abstract}

\maketitle

\section{Introduction}

Competitive team sports are complex systems involving the collective dynamics of players.
Owing to weather conditions, situation-dependent changes of team strategy and other factors, a game can be viewed as a noisy dynamical system, and deterministic predictions for game situation and flow are extremely difficult.
Primarily, the human movement is complex and uncertain, and clear and simple physical laws are not necessarily directly applicable.
However, at least statistically, it is observed that human movement exhibits a certain pattern or predictability~\cite{Barabasi, Buchanan}, and phenomenological models for human movement~\cite{Helbing, Starnini} have been developed.
Similarly, social systems based on interaction of people exhibit statistical regularities~\cite{Castellano, Kobayashi}. 
Effective modeling that can reproduce the properties of real data is a major approach to analyze social and other complex systems, and is valid for sports as well.

Recently, large and precise digital datasets are collected in various sports, which are helpful in providing an effective understanding of the complexity of sports~\cite{Anderson, Kovalchik}.
In particular, football (soccer) is a team sport on which considerable research in physics has been conducted, owing to its worldwide popularity~\cite{Bottenburg, topendsports} and availability of online datasets~\cite{Pettersen, Pappalardo}.
Previously, statistical and phenomenological studies on scores~\cite{Heuer, Everson}, formations~\cite{Narizuka, Kim}, and network structures~\cite{Buldu, Yamamoto} in football matches have been investigated using empirical data analysis, numerical simulation, and theoretical calculations.
Artificial intelligence and machine learning have emerged as additional tools in sports analytics~\cite{Tuyls}, providing new insights into sports data.

In football matches, it is natural to expect ball possession to be an important factor in team success; intuitively, a team holding the ball for a longer period has a greater chance of achieving a goal and win.
However, this issue is controversial~\cite{Anderson, Jones, Wang}.
Some articles have suggested that successful teams garner greater ball possession than unsuccessful teams~\cite{Bradley13} and that ball possession is correlated with match outcome~\cite{Lago10}.
However, other studies have argued that there is no correlation~\cite{Aquino}.
According to Ref.~\cite{Wang}, ``in recent years, the role of possession percentage in the analysis of technical indicators has gradually weakened.''
To further complicate the issue, ball possession has been reported to depend on formation~\cite{Aquino}, match location (home or away)~\cite{Lago11}, and game situation (winning or losing)~\cite{Bradley14}.
Analysis of ball possession is important, not only practically in taking the initiative and creating an effective attack in matches, but also scientifically as a fundamental temporal property in ball sports.

Ball possession percentage is the ratio of total ball possession time for given team to the total possession time of both the given teams and its opponent and is a macroscopic indicator in that it is computed over an entire match.
The macroscopic ball possession can be hierarchically decomposed as follows.
Possession time for a given team in an entire match is the sum of the times the team possessed the ball over all its possession, composed of the possession times of all tis players.
The statistical properties for these two levels of possession time have been previously investigated.
Player possession time was analyzed by Mendes et al.~\cite{Mendes} in South American and European matches, and was found to approximately follow the gamma distribution or its $q$-analog.
Chacoma et al.~\cite{Chacoma} analyzed team possession time as a part of the analysis of a large dataset created by Pappalardo et al.~\cite{Pappalardo} and proposed the inverse Gaussian distribution as a candidate for its probability distribution.

Although player and team possession times have been independently analyzed, the relation between these two levels of possession times has not been explicitly formulated.
As team possession time can be decomposed into the possession times of individual players, it is natural that these two levels of possession times should be closely related to each other.
Nevertheless, there appears to be no known theoretical relation connecting the gamma distribution (for player possession time) and inverse Gaussian distribution (for team possession time).
The current situation is not satisfactory, in which a consistent framework for player and team possession time has not been established.

In this study, the authors propose a theoretical formula relating player and team ball possession times along with the player count of a team ball possession event.
Through the analysis of player-tracking data for matches in the Japan Professional Football League, we confirm that player possession time follows the gamma distribution, qualitatively consistent with the findings of Mendes et al.~\cite{Mendes}.
We find that player count of a team ball possession is effectively approximated by a mixture of two geometric distributions and propose that this distribution is composed of geometric distributions that depend on the spatial location of the possession.
Using the proposed formula and these results, we can obtain an estimate for the distribution of team possession time that is similar to the real distribution.
In this manner, a bottom-up formulation for team possession time from player possession time and player count can be successfully attained.

\section{Dataset}
We analyzed player-tracking and play-by-play data from 95 matches in the top division of the Japan Professional Football League.
Of these, 45 matches were played in 2020 and 50 matches were played in 2021.
The dataset was provided by DataStadium Inc., Japan, and the authors received explicit permission to use the data in this study.
The player-tracking data contain the positional coordinates of each player and the ball every \SI{0.04}{\second}, and the play-by-play data indicate which player touched the ball and when, as well as what action (e.g., a trap, home/away pass, or throw-in) was taken.
By combining these two types of data, we can investigate ball possession events based on spatial location.

According to our experience of watching football matches, goalkeepers (GKs) tend to possess the ball much longer than the other players.
More quantitatively, in our dataset, the player mean possession time, not including GKs, is \SI{2.49}{\second}, whereas the GK mean possession time is \SI{4.61}{\second}.
Because of this exceptionality, we excluded GK possession times from the analysis.
Moreover, we assumed that the chain of ball possession within a team was interrupted even when the GK of the same team held the ball.
The study of player possession time by Mendes et al.~\cite{Mendes} did not considered GK possession; meanwhile, the study of team possession time by Chacoma et al.~\cite{Chacoma} did not specify that GK possession was excluded.

\section{Characteristic quantities}\label{sec3}
In this study, we focus on three quantities connecting ball possession: player possession time, team possession time, and player count of team possession.

The player ball possession event is assumed to begin when a player touches the ball, and continues until another player touches the ball or the match is interrupted (e.g., owing to a foul or ball leaving the field of play).
Player possession time is defined as the time duration of the player ball possession event.
Similarly, we define team possession time as the duration in which players of the team continuously maintain the ball.
Moreover, as noted in the previous section, we consider that the team ball possession event ends when the GK touches the ball.
In particular, we consider the probability densities $f_\mathrm{player}(t)$ and $f_\mathrm{team}(t)$ of player and team possession times, respectively.

Generally, a possession event of a team contains multiple players.
The last characteristic quantity, the player count of a team's possession, represents how many players touch the ball in the possession event.
The player count is increased each time the player in possession of the ball changes if the ball maintained by the same team, including if the same player appears more than once in the chain of possession.
Hence, this number is identical with the number of succeeded passes in a team's possession event plus one.
We use the symbol $n$ for the player count, and define $P(n)$ as the probability or fraction of team possession events with a player count of $n$.

\begin{figure*}[t!]\centering
\includegraphics[scale=0.8]{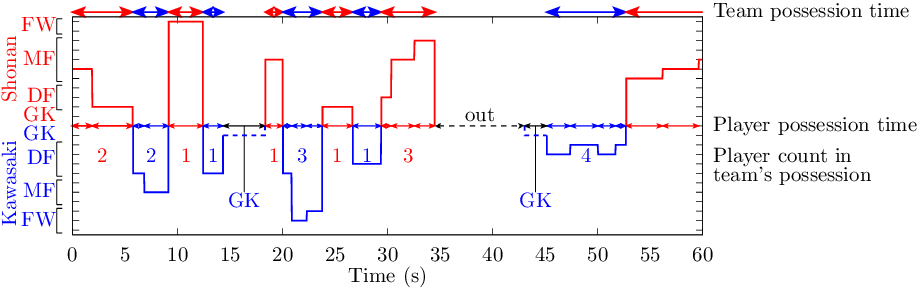}
\caption{
Illustration of the characteristic quantities, considering as an example the flow of ball possession during initial \SI{60}{\second} in Shonan vs Kawasaki (Sept.\ 27, 2020).
Each horizontal line segment represents the interval of ball possession by each player, sorted by positions: GK, DF (defender), MF (midfielder), and FW (forward).
Player possession times are shown with two-way arrows aligned in the middle, team possession times are shown with two-way arrows above the graph, and player counts in team possession are indicated slightly below the middle of the graph.
}
\label{fig1}
\end{figure*}

Figure~\ref{fig1} illustrates the flow of ball possession in a real match and the definitions of our characteristic quantities.
The match was played by Shonan Bellmare (home) and Kawasaki Frontale (away) on Sept.\ 27, 2020.
The two-way arrows appearing in the middle of the graph indicate the possession times, and the arrows above the graph indicate team possession times.
The numbers represent the player counts of team possession events.

\section{Analysis of player possession times}\label{sec4}
\begin{figure}[t!]\centering
\includegraphics[scale=0.8]{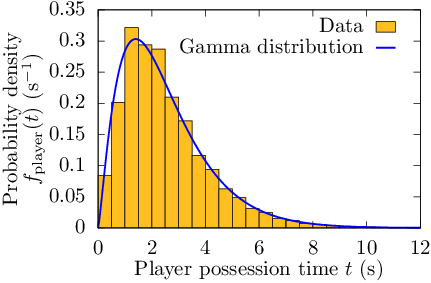}
\caption{
Probability density of player possession time.
The empirical density calculated from all matches in the dataset is shown by the histogram, and the approximated gamma distribution with $\beta=2.29$ and $\tau=\SI{1.09}{\second}$ is represented by the solid curve.
}
\label{fig2}
\end{figure}

Figure~\ref{fig2} shows the probability density $f_\mathrm{player}(t)$ of player possession time.
The histogram represents the empirical density calculated from all 95 matches in the dataset.
The total number of player ball possessions is $1225$ per match.
The histogram displays a unimodal distribution with a mean of \SI{2.49}{\second}.

The solid curve represents the gamma distribution
\[
g(t;\beta, \tau):=\frac{1}{\Gamma(\beta)\tau^\beta}t^{\beta-1}e^{-t/\tau},
\]
where $\Gamma$ is the gamma function.
Using the maximum likelihood estimation~\cite{Ye}, we obtained $\beta=2.29$ and $\tau=\SI{1.09}{\second}$.
(The shape parameter $\beta$ is always dimensionless, and the scale parameter $\tau$ has the dimension of time in this case.)
The overall shape of the empirical distribution is well approximated by the gamma distribution.

The gamma distribution in player possession time was reported by Mendes et al.~\cite{Mendes}, and Fig.~\ref{fig2} is qualitatively consistent with their result.
However, compared to $2.6\le\beta\le4.2$ and $\SI{0.68}{\second}\le\tau\le\SI{1.06}{\second}$ by Mendes et al.~\cite{Mendes}, our result $\beta=2.29$ is small and $\tau=\SI{1.09}{\second}$ is large.
We have no concrete ideas for the reason and meaning of this deviation; however, perhaps the team style of play is captured in $\beta$ and $\tau$.
Mendes et al.~\cite{Mendes} hypothesized that matches played by highly ranked teams have larger $\beta$ and smaller $\tau$, but a comparative study is necessary to validate this hypothesis.

\begin{figure}[t!]\centering
\raisebox{36mm}{(a)}
\includegraphics[scale=0.8]{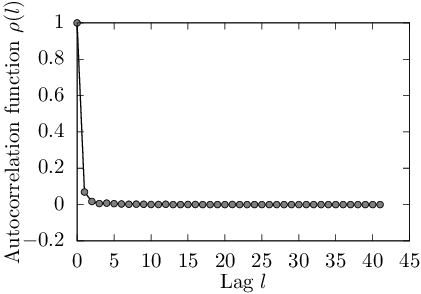}
\hspace{5mm}
\raisebox{36mm}{(b)}
\includegraphics[scale=0.8]{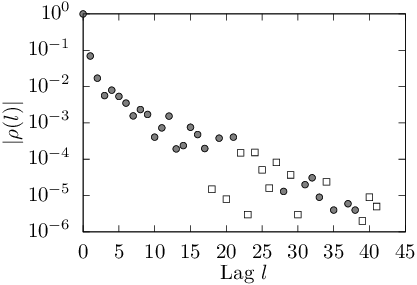}
\caption{
(a) Autocorrelation function $\rho(l)$ of players possession times within a team's possession event.
(b) Semi-log graph for $|\rho(l)|$.
Circles and squares indicate $\rho(l)>0$ and $\rho(l)<0$, respectively.
}
\label{fig3}
\end{figure}

Next, we provide the autocorrelation function of player possession times within a team possession event.
For the total number of team possession events of $S$, with the $s$th event involving $n_s$ players, and the $i$th player possession time in the $s$th event of $t^{(s)}_i$ ($s=1,\ldots,S$ and $i=1,\ldots,n_s$), the autocorrelation function~\cite{Broersen} is defined as
\[
\rho(l)=\frac{\sum_{s=1}^S \sum_{i=1}^{n_s-l} (t^{(s)}_i-\langle T\rangle)(t^{(s)}_{i+l}-\langle T\rangle)}{\sum_{s=1}^S \sum_{i=1}^{n_s} (t^{(s)}_i-\langle T\rangle)^2},
\]
where $\langle T\rangle$ is player mean possession time.
The argument $l$ is called the lag.
Figure~\ref{fig3}(a) shows the numerical result of $\rho(l)$.
The autocorrelation function $\rho(l)$ takes on values near $0$ for $l\ge1$, indicating that successive player possession times have almost no correlation.
This result is used in the theoretical argument presented in Sec.~\ref{sec6}.

To magnify $\rho(l)\approx0$, the semi-log graph of $|\rho(l)|$ is presented in Fig.~\ref{fig3}(b).
Some $l$ give negative correlation; thus, we take the absolute value of $\rho(l)$ in this graph.
The sign of $\rho(l)$, whether $\rho(l)>0$ or $\rho(l)<0$, is distinguished by circles and squares, respectively.
The graph roughly shows a decrease as a whole; however, the decay rate cannot be precisely determined owing to fluctuation.
Some time series data, such as stock prices~\cite{Mantegna} and neuron activity~\cite{Cavanagh}, have been reported to exhibit exponential decay in the autocorrelation function, but player possession time does not display such clear behavior.

\section{Analysis of the player count involved in team possession}
\subsection{Geometric mixture distribution}
\begin{figure}[t!]\centering
\includegraphics[scale=0.8]{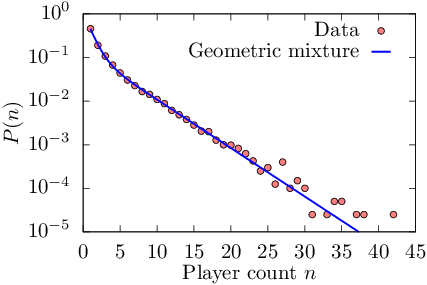}
\caption{
Probability distribution $P(n)$ of player count $n$ on a semi-log scale.
The empirical distribution calculated from all matches is shown as circles, and the optimal geometric mixture distribution~\eqref{eq:geomix} is given by the solid curve.
}
\label{fig4}
\end{figure}

Next we analyze the probability distribution $P(n)$ of player count $n$ in team ball possession.
Circles plotted in Fig.~\ref{fig4} represents the empirical $P(n)$ calculated from all matches in the dataset on a semi-log scale.
The data size (total number of team possession events) is $421.1$ per match, and the mean value of player count $n$ is $2.91$.
As stated in Sec.~\ref{sec3}, the player count is the number of passes plus one, so the mean pass number in a team possession event becomes $2.91-1=1.91\approx2$.Thus, the mean number of possession events of a team in a match is $210.6(=421.1/2)$, each of which contains two passes on average.

The empirical $P(n)$ in Fig.~\ref{fig4} exhibits exponential decay with different slopes in $n\lesssim5$ and $n\gtrsim5$.
Based on this observation, we heuristically apply
\begin{equation}
P(n)=w_1(1-q_1)q_1^{n-1} + w_2(1-q_2)q_2^{n-1} \quad (n=1,2,\cdots).
\label{eq:geomix}
\end{equation}
Each term is composed of the geometric distribution (i.e., the discrete version of the exponential distribution) $P_\mathrm{geo}(n; q)\coloneqq (1-q)q^{n-1}$ ($0<q<1$) times positive constant $w$.
In this paper, we refer to the probability distribution in Eq.~\eqref{eq:geomix} as the \emph{geometric mixture distribution}.
Owing to the normalization of $P(n)$, we consider a constraint for the mixture ratios: $w_1+w_2=1$.
Therefore, the geometric mixture distribution has three independent parameters $q_1$, $q_2$, and either $w_1$ or $w_2$.
Using the expectation-maximization algorithm~\cite{Bishop}, the optimal value is calculated as
\begin{equation}
q_1=0.34,\quad
q_2=0.77,\quad
w_1=0.52,\quad
w_2=0.48.
\label{eq:geomix_param}
\end{equation}
The corresponding geometric mixture distribution is shown by the solid curve in Fig.~\ref{fig4}.

\subsection{Origin of geometric mixture distribution}
This result raises the question why the geometric mixture distribution describes $P(n)$ so well.

We classify team possession events into three types, depending on the location of the ball:
(i) an \emph{own-half} event, in which the ball is always in the own half;
(ii) an \emph{opponent-half} event, in which the ball is always in the opponent half;
and (iii) a \emph{transversal} event, which includes the ball located in both the own and opponent halves.
Note that the own and opponent halves are relative divisions, that is, the own half for one team is the opponent half for the other team and vice versa.
In our dataset, the fractions of own-half, opponent-half, and transversal events are $38.1\%$, $30.5\%$, and $31.4\%$, respectively.

\begin{figure}\centering
\raisebox{36mm}{(a)}
\includegraphics[scale=0.8]{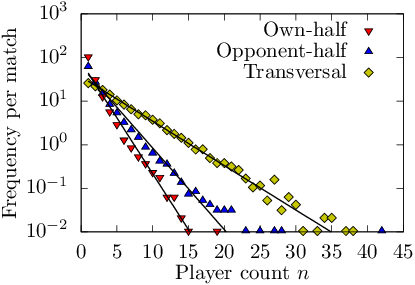}
\hspace{5mm}
\raisebox{36mm}{(b)}
\includegraphics[scale=0.8]{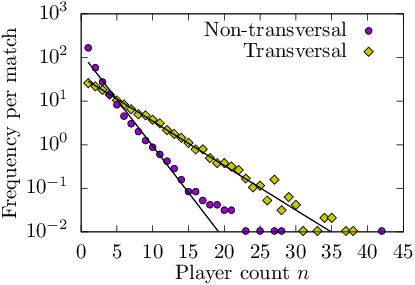}
\caption{
(a) Frequency distribution of player count $n$ for the own-half (inverted triangles), opponent-half (upright triangles), and transversal (diamonds) events.
Three solid lines indicate the optimal geometric distributions.
(b) Circles represent the distribution of non-transversal events, calculated as the sum of the own-half and opponent-half distributions in (a).
Diamonds are the same as in (a).
}
\label{fig5}
\end{figure}

In Fig.~\ref{fig5}(a), inverted triangles, upright triangles, and diamonds represent the frequency distributions of player count $n$ for the own-half, opponent-half, and transversal events, respectively.
Each graph approximately exhibits exponential decay, with the distribution for the transversal events decaying slowly compared to the other two types.
The solid lines indicate the geometric distribution $P_\mathrm{geo}(n; q)=(1-q)q^{n-1}$ with $q_\mathrm{own}=0.55$, $q_\mathrm{opp}=0.64$, and $q_\mathrm{trans}=0.79$.
The distribution of transversal events decays slower than those of own-half and opponent-half events.
There are several reasons for the difference in their decay rates.
First, transversal events require carrying the ball from the own half to the opponent half, which necessarily involves a number of players.
Second, if a player succeeds in getting possession of the ball from an opponent in the own half, he may clear the ball to avoid conceding a goal or may pass the ball to GK to allow his team to rebuild its formation, leading to rapid decay in $P(n)$ for own-half events.
Third, when a player possesses the ball in the opponent half, the opponent players try to dispossess the ball in a desperate attempt to defend their goal, which causes $P(n)$ for opponent-half events to decay rapidly.

The frequency distribution of the own-half or opponent-half events, which we term as \emph{non-transversal} events, is shown as circles in Fig.~\ref{fig5}(b).
This distribution is the simple sum of the own-half and opponent-half distributions in Fig.~\ref{fig5}(a).
The non-transversal distribution can be approximated by a single geometric distribution with $q=0.61$ in $n\le15$, shown as a solid line.

\begin{figure}[t!]\centering
\includegraphics[scale=0.8]{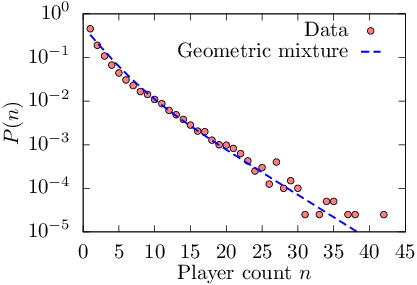}
\caption{
Comparison of the empirical distribution $P(n)$ of player count $n$ (circles) and the geometric mixture distribution estimated by the transversal and non-transversal events: $q_1=0.61$, $q_2=0.79$, $w_1=0.69$, and $w_2=0.31$ (dashed curve).
}
\label{fig6}
\end{figure}

By the definition of the transversal and non-transversal events, the entire distribution $P(n)$ for player count $n$ can be computed by the sum of the transversal and non-transversal distributions.
Corresponding to the slopes of the lines in Fig.~\ref{fig5}(b), we set $q_1=0.61$ (non-transversal) and $q_2=0.79$ (transversal).
As transversal events account for $31.4\%$ as written above, $w_1=1-0.314=0.686$ and $w_2=0.314$.
Figure~\ref{fig6} shows the geometric mixture distribution, with these parameter values represented as a dashed curve, and the empirical distribution $P(n)$ shown as circles.
Although this estimated distribution is inferior to the optimal distribution shown in Fig.~\ref{fig4}, it captures the tendency of the empirical distribution well.
Therefore, we consider that the geometric mixture distribution for $P(n)$ is accounted for by two geometric distributions representing the transversal and non-transversal events.

To be precise, the distribution of the player count $n$ for non-transversal events [circles in Fig.~\ref{fig5}(b)] deviates from the geometric distribution for $n=1$ and $n\gtrsim15$.
We do not believe that division into transversal and non-transversal events is the best and most decisive explanation for the geometric mixture distribution; accordingly, there may be more suitable classifications for team possession events.

\section{Analysis of team possession times}\label{sec6}
\subsection{Relation between player and team possession time}
In this section, we propose a relation between the probability density $f_\mathrm{player}(t)$ of player possession time and $f_\mathrm{team}(t)$ of team possession time, and compare this relation with actual football data.

According to Fig.~\ref{fig3}, player possession times in a team's possession event have almost no correlation; thus, we can reasonably assume that player possession times are independent random variables.
As it is known in probability theory~\cite{Trivedi} that the distribution of the sum of independent random variables is  expressed by the convolution of their individual distributions, the probability density of the sum of the possession times of two players can be written as
\[
(f_\mathrm{player}*f_\mathrm{player})(t)=\int_0^t f_\mathrm{player}(t')f_\mathrm{player}(t-t')dt'.
\]
Similarly, the probability density of the sum of the possession times of $n$ players is given by the $n$-fold convolution
\[
f_\mathrm{player}^{*n}(t)\coloneqq(\overbrace{f_\mathrm{player}*\cdots*f_\mathrm{player}}^n)(t).
\]
Recalling that a team's possession event comprising $n$ players occurs with probability $P(n)$, we obtain
\begin{equation}
f_\mathrm{team}(t)=\sum_{n=1}^\infty P(n)f_\mathrm{player}^{*n}(t).
\label{eq:fteam}
\end{equation}
This is the formula for the distribution of team possession time using the distributions of player possession time and player count.
The right-hand side represents the superposition of $f_\mathrm{player}^{*n}(t)$ with weight $P(n)$ over $n=1,2,\ldots$

In Fig.~\ref{fig2}, we see that $f_\mathrm{player}(t)$ can be approximated by the gamma distribution $g(t; \beta, \tau)$.
It is known that the sum of independent gamma-distributed random variables having identical $\tau$ follows a gamma distribution, which is termed as the reproductive property of the gamma distribution~\cite{Trivedi}.
More specifically, the sum of two independent random variables of probability densities $g(t;\beta_1,\tau)$ and $g(t;\beta_2,\tau)$ has the probability density $g(t;\beta_1+\beta_2,\tau)$.
That is, the probability density of the sum of the possession times of $n$ players becomes $g(t;n\beta,\tau)$.
Meanwhile, the distribution of player count $n$ can be well approximated by the geometric mixture distribution (see Fig.~\ref{fig4}).
Therefore, $f_\mathrm{team}(t)$ for our dataset can be expressed as
\begin{equation}
f_\mathrm{team}(t)=\sum_{n=1}^\infty \sum_{j=1}^2 w_j(1-q_j)q_j^{n-1}g(t;n\beta,\tau).
\label{eq:fteam_data}
\end{equation}

\begin{figure}[t!]\centering
\includegraphics[scale=0.8]{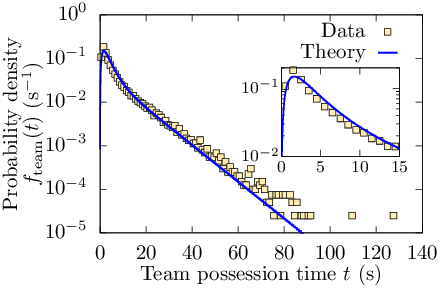}
\caption{
Probability density $f_\mathrm{team}(t)$ of team possession time on a semi-log scale.
Squares are computed from the empirical data of team possession events, and the solid curve is given by Eq.~\eqref{eq:fteam_data}.
The inset shows the enlarged graph in $t<\SI{15}{\second}$.
}
\label{fig7}
\end{figure}

Using parameter values $(\beta, \tau)=(2.29, \SI{1.09}{\second})$ as in Sec.~\ref{sec4} and $(q_1, q_2, w_1, w_2)=(0.34, 0.77, 0.52, 0.48)$ as in Eq.~\eqref{eq:geomix_param}, we can numerically compute $f_\mathrm{team}(t)$ using Eq.~\eqref{eq:fteam_data}.
Figure~\ref{fig7} shows the probability density of team possession time directly computed from the team possession data from all matches, displayed as squares, and the numerical result from Eq.~\eqref{eq:fteam_data} shown as a solid curve.
Our formula~\eqref{eq:fteam_data} yields a good approximation for empirical $f_\mathrm{team}(t)$, except for $t\gtrsim\SI{80}{\second}$ for which possession events occur only infrequently.

The mean of team possession time is \SI{7.24}{\second}.
By using Eq.~\eqref{eq:fteam}, the mean can be calculated as
\[
\int_0^\infty tf_\mathrm{team}(t)dt
=\sum_{n=1}^\infty P(n) \int_0^\infty t f_\mathrm{player}^{*n}(t) dt
=\left(\sum_{n=1}^\infty nP(n)\right)\left(\int_0^\infty t f_\mathrm{player}(t) dt\right),
\]
where
\[
\int_0^\infty t f_\mathrm{player}^{*n}(t)dt=n\int_0^\infty t f_\mathrm{player}(t)dt
\]
by induction.
That is, it is theoretically expected that the mean of team possession time is expressed by the product of the mean player possession time and mean player count of a possession.
In fact, as in previous sections, player possession time and player count in team possession have means of \SI{2.49}{\second} and $2.91$, respectively, and their product $\SI{2.49}{\second}\cdot2.91=\SI{7.25}{\second}$ is very close to the empirical mean \SI{7.24}{\second} of team possession time.

\begin{figure}[t!]\centering
\includegraphics[scale=0.8]{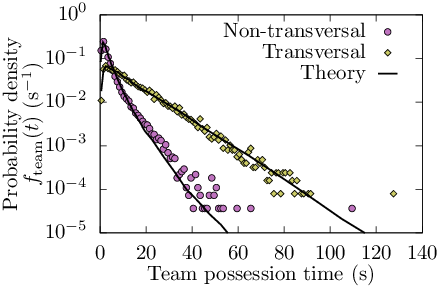}
\caption{
Probability density functions of team possession time for transversal (diamonds) and non-transversal (circle) events.
The solid curves shows the estimated distribution using Eq.~\eqref{eq:fteam} with $P(n)$ and $f_\text{player}(t)$.
}
\label{fig8}
\end{figure}

As shown in the frequency distribution of player count $n$ in Fig.~\ref{fig5}(b), we can split the team possession times into transversal and non-transversal events.
Figure~\ref{fig8} shows the probability density functions of transversal (diamonds) and non-transversal (circles) team possession times.
Transversal events more likely have longer possession times than non-transversal ones, corresponding to the decay rates in distribution $P(n)$ of player count $n$ as shown in Fig.~\ref{fig5}(b).
The solid curves are the estimated functions using Eq.~\eqref{eq:fteam} using empirical distributions of $P(n)$ and $f_\text{player}(t)$.
The proposed formula is consistent with the division of team possession events into transversal and non-transversal ones.

\subsection{Theoretical approximated distribution}
In the previous subsection, we have managed to calculate the mean of team possession time from Eq.~\eqref{eq:fteam}, but this is a fortunate exception.
It appears to be very difficult to derive other theoretical properties of $f_\mathrm{team}(t)$, e.g., its asymptotic behavior, directly from Eq.~\eqref{eq:fteam}, which involves an infinite sum.
Moreover, the exact calculation of the sum in Eq.~\eqref{eq:fteam_data} is likely infeasible, particularly owing to the gamma function $\Gamma(n\beta)$ in $g(t;n\beta,\tau)$.
In this subsection, we introduce several approximations and derive a closed form for $f_\mathrm{team}(t)$.

As written in the previous subsection, the sum of $n$ independent random variables drawn from the gamma distribution with $(\beta, \tau)$ follows the gamma distribution with density $g(t;n\beta, \tau)$.
It is well known that the mean and variance of the gamma distribution $g(t;n\beta,\tau)$ are $n\beta\tau$ and $n\beta\tau^2$, respectively.
Hence, owing to the central limit theorem, we get the following normal approximation:
\begin{equation}
g(t; n\beta, \tau)\simeq\frac{1}{\sqrt{2\pi n\beta\tau^2}}\exp\left(-\frac{(t-n\beta\tau)^2}{2n\beta\tau^2}\right).
\label{eq:normal_approximation}
\end{equation}
Although this normal approximation is only valid for sufficiently large $n$, we shall apply it to all $n=1,2,\ldots$

Next, we use the continuum approximation to replace the discrete sum for $n=1,2,\ldots$ with a continuous integral for $1\le x<\infty$.
The geometric distribution $P_\mathrm{geo}(n; q)$ in the geometric mixture distribution is a discrete distribution, and its continuous counterpart is the exponential distribution
\[
f_\mathrm{exp}(x; \kappa)=\kappa e^{-\kappa (x-1)} \quad(x\ge1).
\]
The usual exponential distribution is defined on $x\ge0$, but we shift the lower bound to $x=1$ in this study.
As we can write $P_\mathrm{geo}(n; q)=(1-q)q^{n-1}=(1-q)q^{-1}\exp(-n\ln(1/q))$, the appropriate relation between $q$ and $\kappa$ is
\[
\kappa = \ln\frac{1}{q} = -\ln q.
\]

Applying these approximations, we have
\begin{align*}
f_\mathrm{team}(t)&\simeq\int_1^\infty \sum_{j=1}^2 w_j\kappa_je^{-\kappa_j(x-1)}\frac{1}{\sqrt{2\pi\beta\tau^2x}}\exp\left(-\frac{(t-x\beta\tau)^2}{2x\beta\tau^2}\right)dx\\
&=\sum_{j=1}^2 w_j\frac{\kappa_j e^{\kappa_j+t/\tau}}{\sqrt{2\pi\beta\tau^2}}\int_1^\infty \frac{1}{\sqrt{x}}\exp\left(-\left(\kappa_j+\frac{\beta}{2}\right)x-\frac{t^2}{2\beta\tau^2x}\right) dx\\
&=\sum_{j=1}^2 w_j\frac{2\kappa_j e^{\kappa_j+t/\tau}}{\sqrt{2\pi\beta\tau^2}}\int_1^\infty \exp\left(-\left(\kappa_j+\frac{\beta}{2}\right)u^2-\frac{t^2}{2\beta\tau^2 u^2}\right)du,
\end{align*}
where we put $u=\sqrt{x}$ in the last equality.
The integral can be calculated using the following formula~\cite{Olver} for $a>0$ and $b\ge0$:
\begin{align*}
\int_1^\infty \exp\left(-a^2u^2-\frac{b^2}{u^2}\right)du
&=\frac{\sqrt{\pi}}{4a}\left[e^{2ab}\erfc(a+b)+e^{-2ab}\erfc(a-b)\right],\\
\end{align*}
where
\[
\erfc(z)\coloneqq\frac{2}{\sqrt{\pi}}\int_z^\infty e^{-x^2}dx
\]
is the complementary error function.
Finally, the approximate form becomes
\begin{align}
f_\mathrm{team}(t)\simeq\sum_{j=1}^2 w_j\frac{\kappa_je^{\kappa_j+t/\tau}}{2\sqrt{\beta(2\kappa_j+\beta)}\tau}
&\left[\exp\left(\sqrt{1+\frac{2\kappa_j}{\beta}}\frac{t}{\tau}\right)\erfc\left(\sqrt{\kappa_j+\frac{\beta}{2}}+\frac{t}{\sqrt{2\beta}\tau}\right)\right.\nonumber\\
&\quad\left.+\exp\left(-\sqrt{1+\frac{2\kappa_j}{\beta}}\frac{t}{\tau}\right)\erfc\left(\sqrt{\kappa_j+\frac{\beta}{2}}-\frac{t}{\sqrt{2\beta}\tau}\right)\right].\label{eq:fteam_approx}
\end{align}

\begin{figure}[t!]\centering
\includegraphics[scale=0.8]{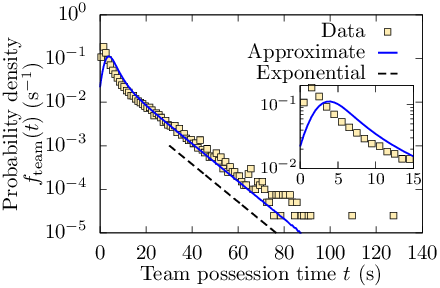}
\caption{
Semi-log graph of the empirical distribution $f_\mathrm{team}(t)$ of team possession time (squares) and its approximate form~\eqref{eq:fteam_approx} (solid curve).
The asymptotic exponential decay~\eqref{eq:fteam_exponential} is shown as the dashed line, shifted downward to avoid overlap with the squares and solid curve.
The inset shows the enlarged graph in $t<\SI{15}{\second}$.
}
\label{fig9}
\end{figure}

As several approximations have been employed to derive Eq.~\eqref{eq:fteam_approx}, we need to check that the approximated $f_\mathrm{team}(t)$ is still consistent with the empirical distribution.
Figure~\ref{fig9} shows this approximated function of $f_\mathrm{team}(t)$ as a solid curve, along with the empirical $f_\mathrm{team}(t)$ shown in squares.
The approximated distribution is deviated from the empirical distribution particularly in $t\lesssim\SI{15}{\second}$ (see the inset of Fig.~\ref{fig9}).
Among the approximations used, the normal approximation of the gamma distribution in Eq.~\eqref{eq:normal_approximation} probably has a large influence on the deviation.
The central limit theorem guarantees that Eq.~\eqref{eq:normal_approximation} is valid for sufficiently large $n\beta$, but its accuracy decreases for small $n$.
A major difference between the normal and gamma distributions is that the normal distribution can take negative values, whereas the gamma distribution cannot.
Therefore, the normal approximation will cause a great deviation for small $t$.
Meanwhile, the approximated distribution decays similarly to the solid curve in Fig.~\ref{fig7} for $t\gtrsim\SI{20}{\second}$.
Hence, Eq.~\eqref{eq:fteam_approx} can be reliably used for the asymptotic form of $f_\mathrm{team}(t)$.

To further simplify Eq.~\eqref{eq:fteam_approx}, we use the asymptotic expansion of the $\erfc$ function~\cite{Olver}:
\[
\erfc(z)\sim\frac{e^{-z^2}}{\sqrt{\pi}z},\quad
\erfc(-z)\sim 2-\frac{e^{-z^2}}{\sqrt{\pi}z}
\]
as $z\to\infty$.
In addition, although $(q_1, w_1)$ and $(q_2, w_2)$ in the geometric mixture distribution~\eqref{eq:geomix} are symmetric by definition, we set $q_1<q_2$ as in Eq.~\eqref{eq:geomix_param}.
We then have $\kappa_1>\kappa_2$, so that $j=2$ in Eq.~\eqref{eq:fteam_approx} becomes dominant for large $t$ and $j=1$ can be neglected.
Finally, we obtain the asymptotic form
\begin{align}
f_\mathrm{team}(t)&\underset{\sim}{\propto}\exp\left(-\left(\sqrt{1+\frac{2\kappa_2}{\beta}}-1\right)\frac{t}{\tau}\right)\nonumber\\
&=\exp\left(-\left(\sqrt{1-\frac{2\ln q_2}{\beta}}-1\right)\frac{t}{\tau}\right).
\label{eq:fteam_exponential}
\end{align}
That is, $f_\mathrm{team}(t)$ exhibits exponential decay asymptotically.
Substituting $\beta=2.29$, $\tau=\SI{1.09}{\second}$, and $q_2=0.77$, the corresponding relaxation time is computed as
\[
\left(\sqrt{1-\frac{2\ln q_2}{\beta}}-1\right)^{-1}\tau\approx\SI{10.1}{\second}.
\]
This exponential slope is shown as a dashed line in Fig.~\ref{fig9} and effectively describes the decay of the empirical distribution.

\section{Discussion}
In this study, the relation between player and team possession times, Eq.~\eqref{eq:fteam}, is established for the first time; these two levels of possession times have been previously analyzed separately.
The proposed result will provide an accurate understanding of the ball possession, which is a fundamental property in ball sports.

We analyze the stochastic properties of player and team ball possession times by summing over the 95 total matches.
A similar analysis for individual match, team and player is a subject for a future study.
For example, there is a possibility that the parameters $(\beta, \tau)$ for each player quantifies the player's style of play and parameters $(q_1, q_2, w_1, w_2)$ for each team quantifies the team's strategy.


While the gamma distribution for player possession time has been reported in the analysis of data from various leagues and championships~\cite{Mendes}, the geometric mixture distribution for player count has not been previously investigated.
Chacoma et al.~\cite{Chacoma} investigated the number of passes in team ball possession, which is our player count minus one.
However, only the graph of the distributions obtained by data analysis and a simplified simulation were provided, neglecting any theoretical discussion.
In addition, because their graph was drawn on a log-log scale, it could not be determined whether the geometric mixture distribution was represented in their data.
Further study of the player count in team possession is required to assess the prevalence of the geometric mixture distribution.

In this study, we have determined that the player count in team possession event follows the geometric mixture distribution, and proposed that this distribution is attributed to the geometric distributions for own-half, opponent-half, and transversal events.
The geometric distribution $P_\mathrm{geo}(n; q)$ for player count $n$ is derived mathematically under the assumption that the ball advances to the next teammate with a constant probability $q$, independent of past events.
This fact may lead to simplified Markovian modeling as in studies on passing networks in football~\cite{Yamamoto18} and rally length in volleyball~\cite{Chacoma22}.
Yet, this simplified assumption does not exactly represent the dynamics in real football matches.
We expect that stochastic modeling based on detailed data analysis allows an evaluation of the validity of the simple geometric distribution, attaching additional meaning to the parameter $q$ and providing an improvement of the geometric distribution.

\begin{figure}[t!]\centering
\includegraphics[scale=0.8]{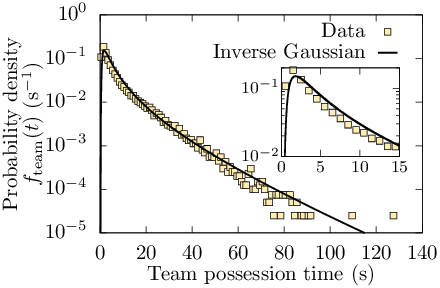}
\hspace{5mm}
\caption{
Semi-log graph for the distribution of team possession time.
The squares and solid curve indicate the empirical distribution and inverse Gaussian distribution, respectively.
The inset shows the enlarged graph in $t<\SI{15}{\second}$.
}
\label{fig10}
\end{figure}

Chacoma et al.~\cite{Chacoma} proposed that team possession time follows the inverse Gaussian distribution, with a probability density of
\[
\sqrt{\frac{\lambda}{2\pi t^3}}\exp\left(-\frac{\lambda(t-\mu)^2}{2\mu^2 t}\right).
\]
The maximum likelihood estimation~\cite{Chhikara} for $\mu$ and $\lambda$ does not adequately fit the entire data; however, a better result was obtained by dropping team possession times less than \SI{1}{\second} as a trial.
Figure~\ref{fig10} shows a semi-log graph for the distribution of team possession time.
The solid curve represents the inverse Gaussian distribution with $\mu=\SI{8.10}{\second}$ and $\lambda=\SI{5.76}{\second}$.
The inverse Gaussian distribution appears to be consistent with the decay of the empirical distribution; however, the evaluation of this result has to consider that short possession times less than \SI{1}{\second}, accounting for $11.3\%$ of the data, were neglected.
If $\mu$ and $\lambda$ of the inverse Gaussian distribution can be expressed in terms of the distributions $f_\mathrm{player}(t)$ and $P(n)$, football data analysis makes considerable progress.


The distribution of team possession time in Eq.~\eqref{eq:fteam_data} and subsequent results~\eqref{eq:fteam_approx} and \eqref{eq:fteam_exponential} depend on the specific form of $f_\mathrm{player}(t)$ and $P(n)$.
Thus, these expressions may vary for other football matches or other ball sport.
In contrast, Eq.~\eqref{eq:fteam} provides a general framework, and is valid for any sports.
We believe that this formula can assist in determining the theoretical properties of player and team possession times.

\section*{Acknowledgments}
The authors are very grateful to DataStadium Inc., Japan for providing the dataset.
This work was partially supported by the Data Centric Science Research Commons Project of the Research Organization of Information and Systems, Japan.
K.Y. was supported by a Grant-in-Aid for Scientific Research (C) (23K03264) from the Japan Society for the Promotion of Science (JSPS), and T.N. was supported by a Grant-in-Aid for Early-Career Scientists (23K16729) from JSPS.

\end{document}